\documentclass{article}
\usepackage{amsfonts}
\usepackage{amsmath}

\input{tcilatex}

\begin{document}

\title{Theory of Critical Temperature Adiabatic Change for Ideal Gas
Bose-Einstein Condensation in Optical Lattices}
\author{G.A. Muradyan, \ A. Zh. Muradyan \\
Department of Physics, Yerevan State University, \\
1 Alex Manookian 375025, Yerevan Armenia\\
e-mail: gmurad@ysu.am}
\maketitle

\begin{abstract}
We present a scheme of analytical calculations determining the critical
temperature and the number of condensed atoms of ideal gas Bose-Einstein
condensation in external potentials with 1D, 2D or 3D periodicity. \ In
particular we show that the width of the lowest energy band appears as the
main parameter determining the critical temperature of condensation. Is
obtained a very simple, proportional to 1/3 degree, regularity for this
dependence. The fundamental role of tunneling in physics of condensate
establishment is underscored.
\end{abstract}

\section{\protect\bigskip Introduction}

Degenerate Bose gases provide an excellent ground for the theoretical study
of quantum fluids since their diluteness makes possible first-principles
approaches\cite{1}. Thanks to today's atomic physics powerful experimental
techniques their properties can be studied quantitatively through a wide
range of temperature and densities. In the last few years Bose-Einstein
condensates (BEC) and Fermi gases in optical lattices have been an extremely
active area of research\cite{2}. Periodic potentials have been used to
examine the transport of Bose-condensed samples \cite{3}, to investigate
effects correlated with the physics of strongly correlated many-body systems 
\cite{4}. These systems can be possibly implemented in quantum information
processing \cite{5} and there are proposals on how to build quantum gates 
\cite{6},\cite{7} and qubit buses \cite{8} for information exchange.

The presence of trapping potentials has a big impact on characteristics of
the condensate, and in particular on the value of temperature at which the
gas passes from normal into BEC state and a macroscopic order coherence
starts to form. \ The predicted increase of critical temperature in a trap
played a big role in obtaining the BEC state with the help of laser cooling
and evaporative cooling techniques. \ However presence of the external
potential doesn't always bring to the increase of critical temperature of
condensation ($T_{C}$). \ In \cite{9} was shown that if the external
potential is periodic, $T_{C}$ decreases and for asymptotically deep
potentials tends to $0$. The decrease of critical temperature for relatively
deep potentials, \ created as off-resonant standing waves, was observed
experimentally in \cite{10}. \ We also gave a descriptive physical
interpretation for the dependence of $T_{C}$ on external trapping potential
parameters \cite{11}, basing on behavior of the distance between the
low-laying energy levels of translational atom. For strictly periodic
potentials this role is naturally played by widths of the low energy bands,
distances between them and most of all the width of the lowest energy band.

In this paper, using some nondrastic mathematical approximations, we derive
an elementary analytic expression for critical temperature $T_{C\text{ }}$
as a function of optical lattice\ parameters. The potential periodicity is
assumed in one, two or three independent directions. The motion for the left
directions in first two cases is assumed to be free.\ The analytical result
justifies, in particular, our physical reasonings brought in \cite{11} about
the $T_{C\text{ }}$ behavior relative to lattice depth adiabatic changes,
and concretizes the form of that behavior. The obtained results, we hope,
will help in experimentally more accurately determining the range of
parameters, where the ultracold Bose gases behave as ideal.

\section{The statistical problem of ideal Bose-gas in periodic potentials}

The impact of laser radiation fields on translational motion of atoms is
presented as a momentum exchange in photon absorption and emission
processes. \ This means that, in general, the atom (molecule) translational
state evolution is connected with the evolution of internal states, leading
to notion of potential for each energy level. In the case of large resonance
detunings, however, it becomes possible to introduce the idea of potential
for the center of mass, as for unstructured particles.\ Just such a
situation will be assumed later, taking the laser field in form of standing
waves creating the periodic potential. The coordinate space in our
calculations is ordinary, three dimensional, while the periodic potential is
present on one, two, or in all three directions.

The principal grand canonical relation, relating the chemical potential $\mu 
$ with the number of atoms $N$ in a system of volume $V$\cite{12}, in case
of interest takes the form%
\begin{equation}
\frac{k^{3}V}{\pi ^{3}}\int_{-\infty }^{+\infty }dP_{X}\int_{-\infty
}^{+\infty }dP_{Y}\int_{-\infty }^{+\infty }dP_{Z}\frac{1}{\exp [\frac{%
\varepsilon _{X}(P_{X})+\varepsilon _{Y}(P_{Y})+\varepsilon _{Z}(P_{Z})-\mu 
}{\kappa _{B}T}]-1}=N.  \label{1}
\end{equation}%
Here $P_{j}(j=X,Y,Z)$ is the atomic momentum for free motion directions and
is the quasimomentum for the periodic potential directions. \ Momentum
(quasimomentum) is scaled by the recoil momentum $2\hbar k$, where $k$ is
the wavevector of the counterpropagating waves forming the standing wave.
The energies $\varepsilon _{j}(P_{j})$, as well as the chemical potential $%
\mu $ and the thermal energy $\kappa _{B}T$, are scaled in recoil energy
units $E_{R}=(2\hbar k)^{2}/2M$, corresponding to momentum value $2\hbar k$.
Formulae (1) assumes an extended form for the dispersion relation between
energy and quasimomentum.

\subsection{1D periodic potential}

After taking the elementary integrals over two free directions in (1), we get%
\begin{equation}
-2\frac{k^{3}V\kappa _{B}T}{\pi ^{2}}\int_{0}^{+\infty }dP\ln \left[ 1-\exp %
\left[ \frac{\mu -\varepsilon (P)}{\kappa _{B}T}\right] \right] =N,
\label{2}
\end{equation}%
where the integration over the total range of quasimomentum is replaced by
two half-range integrals - from zero to infinity. To proceed, it is
convenient to divide the whole range of integration into a sequence of
Brillouin zones and expand the integrand logarithm into convergent Taylor
series. \ Then the above relation takes the following form:%
\begin{equation}
2\frac{k^{3}V\kappa _{B}T}{\pi ^{2}}\sum_{n=1}^{\infty }\frac{1}{n}\exp [n%
\frac{\mu }{\kappa _{B}T}]\sum_{m=1}^{\infty }\int_{(m-1)/2}^{m/2}dP\exp %
\left[ -n\frac{\varepsilon (P)}{\kappa _{B}T}\right] =N.  \label{3}
\end{equation}%
Now let's temporarily concentrate on the integral term and take a dispersion
relation, for example, for a biparabolic form of the periodic potential,
introduced in \cite{11}:%
\begin{equation}
\cos (2\pi P)=1+2G_{11}(\varepsilon )G_{22}(\varepsilon ),  \label{4}
\end{equation}%
(the explicit expressions of $G_{11}(\varepsilon )$ and $G_{22}(\varepsilon
) $ are not necessary for our later presentation).

As is seen from Fig.1, the dependence of energy $\varepsilon ~$on$\ $%
quasimomentum $P$, calculated by relation (4), is almost linear in frame of
the first energy band and this linearity isn't rapidly lost for higher
energy bands. Such a behavior prompts us to introduce a new, main in context
of this paper, approximation taking the $\varepsilon (P)$ -dependence inside
each energy band as linear. Here we come from the fact, that in assumed
thermal equilibrium state of the Bose gas the population and respective
contribution of upper energy bands into the left-side value of (3) decreases
quite rapidly.\ 

After denoting boundary energy values of the m-$th$ zone by $\varepsilon
_{\min }^{(m)}$ and $\varepsilon _{\max }^{(m)}$ and performing the
mentioned linearization, the dispersion relation (4) will have the form%
\begin{equation}
\varepsilon (P)=\frac{\varepsilon _{\max }^{(m)}+\varepsilon _{\min }^{(m)}}{%
2}+(-1)^{m}\frac{\varepsilon _{\max }^{(m)}-\varepsilon _{\min }^{(m)}}{2}%
\cos (2\pi P),  \label{5}
\end{equation}%
and the integral in (3) is expressed by zero order modified Bessel function $%
I_{0}(x)$ with an exponential factor. \ As a result we obtain the following,
more simple form for the main statistical relation:%
\begin{equation}
\frac{k^{3}V\kappa _{B}T}{\pi ^{2}}\sum_{m=1}^{\infty }\left\{
\sum_{n=1}^{\infty }\frac{1}{n}\exp \left( n\frac{\mu -\bar{\varepsilon}%
^{(m)}}{\kappa _{B}T}\right) I_{0}\left( n\frac{\delta ^{(m)}}{2\kappa _{B}T}%
\right) \right\} =N,  \label{6}
\end{equation}%
where $\bar{\varepsilon}^{(m)}=(\varepsilon _{\max }^{(m)}+\varepsilon
_{\min }^{(m)})/2$ is the mean energy in the m-$th$ band and the parameter $%
\delta ^{(m)}=\varepsilon _{\max }^{(m)}-\varepsilon _{\min }^{(m)}$ in the
argument of modified Bessel function is the width of that band. The
contribution of each energy band in this relation is now determined by the
expression in curve brackets and presents a single-variable convergent
series, a very convenient form for numerical calculations. To be convinced
in convergent nature of the mentioned series, one would address to
asymptotic formulae of the modified Bessel function for great values of
argument:%
\begin{equation}
I_{0}\left( n\frac{\delta ^{(m)}}{2\kappa _{B}T}\right) \approx \frac{\sqrt{%
2\kappa _{B}T}}{\sqrt{2\pi n\delta ^{(m)}}}\exp \left( n\frac{\delta ^{(m)}}{%
2\kappa _{B}T}\right) .  \label{7}
\end{equation}%
The asymptotic behavior of the mentioned sum is then determined by
expression 
\begin{equation}
\sum_{large\text{ }n}\frac{1}{n^{3/2}}\exp \left[ -\frac{n}{\kappa _{B}T}%
(\varepsilon _{\min }^{(m)}-\mu )\right] ,  \label{8}
\end{equation}%
which evidently converges for any $\mu \leq \varepsilon _{\min }^{(m)}$, as
usual for bosonic systems.

Thus, from mathematical viewpoint relation (6) is reasonably defined and is
convenient for calculating the functional dependence of $\mu $ on $T$ \ with
any in advance prescribed accuracy. \ In particular the critical temperature
will be decided by substituting $\mu =\varepsilon _{\min }^{(1)}:$%
\begin{equation}
\frac{k^{3}V\kappa _{B}T}{\pi ^{2}}\sum_{m=1}^{\infty }\left(
\sum_{n=1}^{\infty }\frac{1}{n}\exp \left( -n\frac{\bar{\varepsilon}%
^{(m)}-\varepsilon _{\min }^{(1)}}{\kappa _{B}T_{C}}\right) I_{0}\left( n%
\frac{\delta ^{(m)}}{2\kappa _{B}T_{C}}\right) \right) =N.  \label{9}
\end{equation}

From general principles of the Bose -Einstein condensation theory directly
follows that for $T<T_{C}$ in (6) we should interpret $N$ as the number of
noncondensed atoms($N_{nc}$) and not as the total number, simultaneously
taking $\mu =\varepsilon _{\min }^{(1)}$ which corresponds to the critical
temperature. The number of condensed atoms is determined by the
complementary relation $N_{c}=N-N_{nc}.$

Now let's go back to formulae (7) and use it not for showing the convergence
of the series, but for getting a new, much more simple approximate form of
problem solution. Really\textbf{,} the formulae (7) can be used for small
values of $n$ too, if first of all $\kappa _{B}T<\delta ^{(1)}/2,$ that is
if atom's thermal energy is appreciably smaller than the half-width of the
first, most narrow energy band (for critical and lower temperatures this
condition is well satisfied in today's experiments). Nevertheless,in
general, the fulfillment of this condition is not mandatory and the use of
(7) for the first addends too would be allowed approximation when the impact
of these first members is small. \ Without going into details we will just
mention that for the above mentioned replacement is enough to satisfy the
condition 
\begin{equation}
\frac{\pi ^{2}N}{k^{3}V\kappa _{B}T}>\frac{2\kappa _{B}T}{\delta ^{(m)}}.
\label{10}
\end{equation}%
After substituting (7) into (6) the sum over $n$ can be calculated and
expressed by confluent hypergeometric function $\Phi (a,3/2;1)$. Then the
main statistical relation takes the form%
\begin{equation}
\frac{k^{3}V}{\pi ^{5/2}}(\kappa _{B}T)^{3/2}\sum_{m=1}^{\infty }\frac{1}{%
\delta ^{(m)}}\exp \left( -\frac{\varepsilon _{\min }^{(m)}-\mu }{\kappa
_{B}T}\right) \Phi (e^{-(\varepsilon _{\min }^{(m)}-\mu )/\kappa
_{B}T},3/2;1)=N,  \label{11}
\end{equation}%
from which the critical temperature is decided as before, substituting $\mu
=\varepsilon _{\min }^{(1)}$:

\begin{equation}
\frac{k^{3}V}{\pi ^{5/2}}(\kappa _{B}T_{c})^{3/2}\left\{ \frac{\Phi (1,3/2;1)%
}{\delta ^{(1)}}+\sum_{m=2}^{\infty }\frac{1}{\delta ^{(m)}}\exp \left( -%
\frac{\varepsilon _{\min }^{(m)}-\varepsilon _{\min }^{(1)}}{\kappa _{B}T_{c}%
}\right) \Phi (e^{-(\varepsilon _{\min }^{(m)}-\varepsilon _{\min
}^{(1)})/\kappa _{B}T_{c}},3/2;1)\right\} =N.  \tag{12}
\end{equation}%
First term in curve brackets stands for the number of atoms in the first
energy band. The number of atoms in any, higher laying energy band is
determined by the respective addend of the series: $m=2$ for the second
(first excited) one, and so on.

Often in experiments with optical lattices we can limit ourselves with the
first energy band. For this cases we get elementary relation and for making
its appearance more contensive, it is appropriate to introduce a notion of
the first energy band for free (when the energy gaps tend to zero) Bose gas.
If we denote this width by $\delta _{0}^{(1)}$(in normalized units it is
equal to 1/4), and the critical temperature of a free ideal gas \cite{12} by 
$T_{C0}$ , than we arrive at the following final form:%
\begin{equation}
T_{C}=\left( \frac{\pi }{2}\right) ^{2/3}\left[ \frac{\delta ^{(1)}}{\delta
_{0}^{(1)}}\right] ^{1/3}T_{C~0}.  \label{13}
\end{equation}

This formulae is one of the main results of the present paper. The width $%
\delta ^{(1)}$\textit{\ }is the only parameter depending on external
potential.\ \ The deepening of the periodic potential, as is well known,
compresses this width and, cosequently, decreases $T_{C},$ the critical
temperature of condensation (such a result, based on numerical calculations
and physical reasonings, we have presented in \cite{11}). The formulae shows
that this decrease happens by a very simple law: proportional to 1/3 degree
of the first energy band width.

The approximate character of formulae (13) is seen in fact, that in the
limit of free gas ($\delta ^{(1)}\rightarrow \delta _{0}^{(1)}$) the
critical temperature $T_{C}$ tends not to $T_{C~0}$ but to $\left( \pi
/2\right) ^{2/3}\approx 1.35$ times higher value. \ For not very deep
potentials (in $E_{r}$ units smaller than one) the main approximation done
here is the substitution of dispersion curves with straight lines. The
deepening of the potential (decreasing $\delta ^{(m)}$) straightens the
dispersion curves, starting from the first energy band, and consequently
softens the role of this approximation. But now grows the role of using the
asymptotic formulae (7), as for the first members of series the condition $%
n\delta ^{(m)}/2\kappa _{B}T>1$ fails. The single band approximation for
ultracold gases has a minor impact whatever the case.

\subsection{2D and 3D periodic potentials}

Suppose that 2D (3D) periodic potential is a sum of two (three) periodic
potentials, each one of which is periodic only in one direction. \ The
calculation scheme doesn't undergo any qualitative changes and we will write
the final results straight away. Instead of (6), we come to relation%
\begin{equation*}
\frac{k^{3}V\kappa _{B}T}{\pi ^{5/2}}\sum_{m,m^{/}=1}^{\infty }\left( 
\begin{array}{c}
\sum_{n=1}^{\infty }\frac{1}{\sqrt{n}}\exp \left( n\frac{\mu -\bar{%
\varepsilon}^{(m,X)}-\bar{\varepsilon}^{(m^{/},Y)}}{\kappa _{B}T}\right)
\times \\ 
I_{0}\left( n\frac{\delta ^{(m,X)}}{2\kappa _{B}T}\right) I_{0}\left( n\frac{%
\delta ^{(m^{/},Y)}}{2\kappa _{B}T}\right)%
\end{array}%
\right) =N
\end{equation*}%
for 2D periodicity and%
\begin{equation*}
\frac{k^{3}V\kappa _{B}T}{\pi ^{3}}\sum_{m,m^{/},m^{//}=1}^{\infty }\left( 
\begin{array}{c}
\sum_{n=1}^{\infty }\exp \left( n\frac{\mu -\bar{\varepsilon}^{(m,X)}-\bar{%
\varepsilon}^{(m^{/},Y)}--\bar{\varepsilon}^{(m^{//},Z)}}{\kappa _{B}T}%
\right) \times \\ 
I_{0}\left( n\frac{\delta ^{(m,X)}}{2\kappa _{B}T}\right) I_{0}\left( n\frac{%
\delta ^{(m^{/},Y)}}{2\kappa _{B}T}\right) I_{0}\left( n\frac{\delta
^{(m^{//},Z)}}{2\kappa _{B}T}\right)%
\end{array}%
\right) =N
\end{equation*}%
for 3D periodicity, where the new notations are the full analogies of the\
1D case. After applying (7) these relations are simplified and take the forms%
\begin{equation*}
\frac{k^{3}V}{\pi ^{7/2}}(\kappa _{B}T)^{3/2}\sum_{m,m^{/}=1}^{\infty
}\left( 
\begin{array}{c}
\frac{1}{\sqrt{\delta ^{(m,X)}\delta ^{(m^{/},Y)}}}\exp \left( -\frac{%
\varepsilon _{\min }^{(m,X)}-\varepsilon _{\min }^{(m^{/},Y)}-\mu }{\kappa
_{B}T}\right) \times \\ 
\Phi \left( e^{-(\varepsilon _{\min }^{(m,X)}-\varepsilon _{\min
}^{(m^{/},Y)}-\mu )/\kappa _{B}T},\frac{3}{2};1\right)%
\end{array}%
\right) =N
\end{equation*}%
and%
\begin{equation*}
\frac{k^{3}V}{\pi ^{9/2}}(\kappa _{B}T)^{3/2}\sum_{m,m^{/},m^{//}=1}^{\infty
}\left( 
\begin{array}{c}
\frac{1}{\sqrt{\delta ^{(m,X)}\delta ^{(m^{/},Y)\delta ^{(m^{//},Z)}}}}\exp
\left( -\frac{\varepsilon _{\min }^{(m,X)}-\varepsilon _{\min
}^{(m^{/},Y)}+\varepsilon _{\min }^{(m^{//},Z)}-\mu }{\kappa _{B}T}\right)
\times \\ 
\Phi \left( e^{-(\varepsilon _{\min }^{(m,X)}-\varepsilon _{\min
}^{(m^{/},Y)}+\varepsilon _{\min }^{(m^{//},Z)}-\mu )/\kappa _{B}T},\frac{3}{%
2};1\right)%
\end{array}%
\right) =N
\end{equation*}%
correspondingly. For critical temperatures we get just as simple relations
as was (13):%
\begin{equation}
T_{C}\approx \left( \frac{\pi }{2}\right) ^{4/3}\left[ \frac{\delta ^{(1,X)}%
}{\delta _{0}^{(1)}}\right] ^{1/3}\left[ \frac{\delta ^{(1,Y)}}{\delta
_{0}^{(1)}}\right] ^{1/3}T_{C\ 0}  \label{14}
\end{equation}%
for potentials with 2D and%
\begin{equation}
T_{C}\approx \left( \frac{\pi }{2}\right) ^{2}\left[ \frac{\delta ^{(1,X)}}{%
\delta _{0}^{(1)}}\right] ^{1/3}\left[ \frac{\delta ^{(1,Y)}}{\delta
_{0}^{(1)}}\right] ^{1/3}\left[ \frac{\delta ^{(1,Z)}}{\delta _{0}^{(1)}}%
\right] ^{1/3}T_{C\ 0}  \label{15}
\end{equation}%
with 3D periodicity.

\section{Conclusions}

Formulas (12), (13), (14) and (15) represent the main results of this paper.
They show that with good approximation the width of the first energy band is
the only parameter determining the ideal gas Bose-Einstein condensation
critical temperature in a field of periodic potentials. \ \ From the Bloch
theory of periodic potentials is well known that the deepening of the
potential rapidly narrows energy bands and especially the first one.
Therefore even the 1/3 degree proportionality in 1D case \ will be enough
for decreasing the critical temperature twice for $2E_{R}-3E_{R}$ potential
depths. \ The comparison of the above mentioned formulas also shows that in
a single-band approximation each direction of periodicity acts as though
independent and by that speeds up the critical temperature dependence on the
potential depth when the periodicity passes from 1D to 2D and 3D
correspondingly.

The band structure of energy spectrum assums the existence of Bloch-type
wavefunctions, the modulus of which is periodically spread over the whole
potential. For the states which are of interest, that is with energies
smaller than the potential height, the only mechanism of the Bloch state
realization is the quantum tunneling through the potential barriers. By this
we conclude that the phenomena of tunneling has an exceptional role for the
obtained BEC critical temperature behavior regularity. \ If we exclude the
possibility of tunneling for quantum particles (in our case of atom), then
each low-energy atom would've stay trapped and localized only in one
well-type region of the potential, as for example in harmonic potential
case. The atom translational motion energy spectrum would be discrete and
the energy levels would move away as the potential depth increases. Then it
would bring to an increase of the critical temperature, a result well known
for isolated wells and just opposite to the above obtained ones.

\section{Acknowledgments}

One of the authors (A. Zh. M.) thanks H. Metcalf, M. Kasevich and members of
their groups for active discussions. This work was supported by Armenian
State Funding Program 0126.

New-York 1975); K.Huang, \textit{Statistical Mechanics}, (John Wiley\&Sons,
New York 1963).

\FRAME{ftbpFU}{3.3736in}{1.9804in}{0pt}{\Qcb{The dependence of the right
hand side(RHS) of (4) from normalized energy $\protect\varepsilon $. The
optical lattice depth is taken $1E_{R}$. The bold intervals of the curve
correspond to allowed energy bands.}}{\Qlb{Fig.1}}{fig.1.eps}{\special%
{language "Scientific Word";type "GRAPHIC";maintain-aspect-ratio
TRUE;display "USEDEF";valid_file "F";width 3.3736in;height 1.9804in;depth
0pt;original-width 3.3451in;original-height 1.9519in;cropleft "0";croptop
"1";cropright "1";cropbottom "0";filename 'E:/Documents and
Settings/ports/My 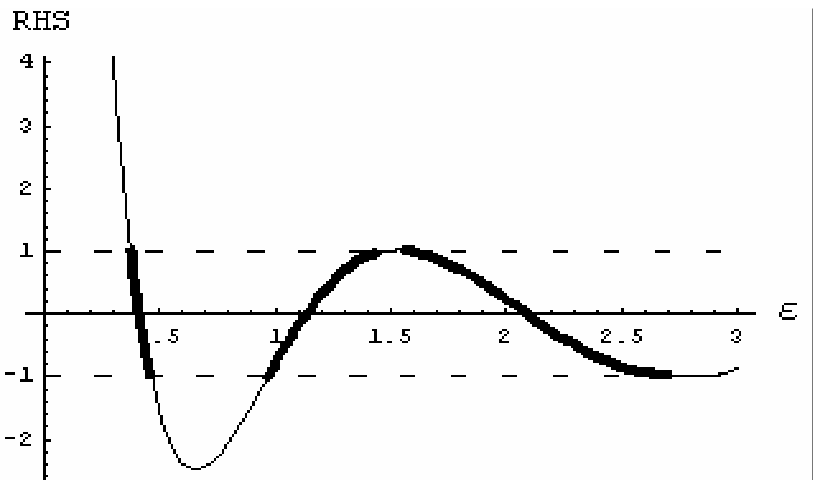';file-properties "XNPEU";}}


\bigskip

\bigskip

\bigskip

\begin{thebibliography}{99}
\bibitem{1} \bigskip "\textit{Bose-Einstein condensation in atomic gases}%
\textquotedblright ; edited by M. Inguscio, S. Stringari and C. Wieman, (IOS
Press, Amsterdam 1999); A.J. Leggett, Rev. Mod. Phys. \textbf{73}, 307
(2001), J.R. Anglin, and W. Ketterle, Nature \textbf{416,} 211 (2002).

\bibitem{2} M. Greiner, et al., Phys. Rev. Lett. \textbf{87}, 160405 (2001);
M. Greiner, et al., Nature \textbf{419}, 51 (2002); C. Orzel, et al.,
Science \textbf{293}, 2386 (2001); F. S. Cataliotti, et al., Science \textbf{%
293}, 843 (2001); T. St\"{o}ferle, et al., Phys. Rev. Lett. \textbf{96},
030401 (2006); M. K\"{o}hl, et al., Phys. Rev. Lett. \textbf{94}, 080403
(2005); M. K\"{o}hl, Phys. Rev. A \textbf{73}, 031601(R) (2006); C. Schori,
et al., Phys. Rev. Lett. \textbf{93}, 240402 (2004); S. Ospelkaus, et al.,
Phys. Rev. Lett. \textbf{96,} 180403 (2006).

\bibitem{3} B. P. Anderson and M. A. Kasevich, Science \textbf{282}, 1686
(1998).

\bibitem{4} M. Greiner, et al., Nature \textbf{415}, 39 (2002).

\bibitem{5} D. Jaksch and P. Zoller, arXive: \textit{cond-mat}/0410614.

\bibitem{6} O. Mandel et al., Nature \textbf{425}, 937 (2003).

\bibitem{7} see D. Vager, B. Segev and Y. Band, arXive:\textit{quant-ph}%
/0505199 and references therein.

\bibitem{8} G. Brennen et al, Phys. Rev. A\textbf{\ 67}, 050302 (2003).

\bibitem{9} A.Zh. Muradyan, H.L. Haroutyunyan, Izvestiya NAN Armenii, Fizika 
\textbf{35}, 3 (2000).

\bibitem{10} S. Burger et al, Europhys. Lett. \textbf{57}, 1 (2002).

\bibitem{11} A.Zh. Muradyan and G.A. Muradyan, arXive: \textit{cond-mat}%
/0302108.

\bibitem{12} R. Balescu, \textit{Equilibrium and nonequilibrium statistical
mechanis}, (Wiley-Interscience, 
\end{thebibliography}
\end{document}